# Seek These High-Confidence Biosignatures in Mid-Latitude Martian Ice

Christopher Temby[1,2], Jan Spacek[1,3] (info@alfamars.org)
[1]Agnostic Life Finding Association (ALFA); [2]University of Florida (UF), [3]Foundation for Applied Molecular Evolution (FfAME), Alachua, FL, USA

To ensure that the first mission designed to seek signs of extant life since 1976 is able to produce an unambiguous biological interpretation, the SFL-SAG is tasked with identifying the most high-confidence, agnostic biosignatures which are targetable, detectable, and measurable in Martian subsurface mid-latitude ice. To aid in this effort, this white paper highlights three examples of target materials or phenomena, along with associated instrument concepts, which the SFL-SAG shall prioritize in its efforts to define the appropriate astrobiological strategy.

First, to enable a confident detection of alien life, particularly in an oligotrophic and low-biomass environment, the mission shall produce unprecedentedly large sample sizes to adequately assess an ultra-low-biomass biosphere, accounting for the analytical instrument sensitivity, expected population densities, and instrument detection confidence requirements. Given the extremely-low population density expected on Mars, concentration of target materials prior to analysis will be required for these large sample sizes. Furthermore, to avoid ambiguous interpretations, the mission architecture may require mobile surface operations (drilling-capable rover [1, 2]) to prospect enough of the Martian subsurface. Significant technology development will be needed to meet these requirements.

Additionally, to minimize false negative and false positive results, the astrobiological payload must target, detect, and measure high-confidence biosignatures that are agnostic to the natural history and biochemistry of the alien biosphere. As highlighted by the SFL-SAG's initial findings, these would include molecules conferring structure or function, metabolism, and/or agnostic high-molecular-complexity polymers. Of these, polyelectrolyte informational biopolymers are an ideal target material [3, 4]. Polyelectrolytes are high-confidence, agnostic biosignatures that confer function, and their detection and sequencing minimizes false positive and false negative interpretations. Furthermore, since the conferred function is "information inheritance," one can infer the existence of Darwinian evolution.

The "Ladder of Life Detection" paper by Neveu et al (2018) [5], authored by NASA HQ staff, endorses a polyelectrolyte polymer as a worthy biosignature for a mission searching for alien life. However, concerns are raised that polyelectrolyte "detectability may be hampered by dilution" and their "survivability… is limited by hydrolysis." The former concern is easily solved by utilizing electrodialysis to concentrate polyelectrolytes from a dilute solution [4]. Addressing the latter concern, the long-term instability of polyelectrolytes in water is actually a benefit when seeking extant life: if polyelectrolytes *are* detected, their long-term instability means that extant life *must* be replenishing them. Additionally, extremely large sample volumes along with nanopore sequencing and characterizations will reduce the chance of false positives due to forward contamination.

Accordingly, the Agnostic Life Finder (ALF) has been designed to isolate, desalt, and concentrate sparse polyelectrolytes from large volumes of water. To enable this, ALF uses continuous electrodialysis with porous membranes to isolate concentrate polyelectrolytes (DNA, RNA, or alien informational biopolymers) from water, based on size and charge. Polyelectrolytes move in an electric field, and are bigger than inorganic electrolytes (salts), allowing ALF to concentrate them by electromigration in water through size-exclusionary membranes. ALF can achieve an arbitrarily low limit of detection for informational biopolymers, where the sensitivity is limited only by the sample volume that ALF inspects [6].

Following ALF's preconcentration step, known genetic material recovered will be sequenced using biological nanopores. Unknown polyelectrolytes will be analyzed using solid-state nanopores and/or fragmentation mass spectrometry to determine whether the heteropolymers are composed of a limited set of building blocks. The ALF instrument is currently TRL-4, following support from NASA's NIAC program, with development on track for TRL-8 before 2030. The ALF system comprises several subsystems: (1) a preconcentrator, (2) a desalting unit, (3) a sonicator, (4) the ALF stack, (5) a polyelectrolyte concentrator, and (6) analyzers.

The Agnostic Life Finding Association (ALFA) and the University of Florida (UF) support the development of instrument concepts that enable the analyses/measurements of agnostic biosignatures. In addition to the aforementioned ALF system, ALFA supports the development of the Integrated Miniature Polarimeter and Spectrograph (IMPS) instrument. IMPS is a bench-top instrument that enables the non-destructive chiroptical spectroscopy of macromolecular biological homochirality. It is yet to be determined what levels of preconcentration are needed to provide adequate in-situ sensitivity for chiroptical detection. Chiroptical measurements can be done for many sample types, but would be particularly useful as an orthogonal measurement of polyelectrolyte biopolymers, in addition to ALF's aforementioned analyses.

ALFA advises NASA's SFL-SAG to recognize that an unambiguous interpretation of biology can be concluded following the sequencing of known genetic polymers (DNA and RNA), and for unknown polymers, the same interpretation can be concluded following analysis with solid state nanopore and potentially with the chiroptical analysis. However, recognizing the utility of further context and orthogonal detections of additional target materials, ALFA agrees with the SFL-SAG's intention of identifying several biosignatures or phenomena and measurements, provided their sensitivity is sufficient to determine at least single (alien) cell per gram of Martian sample.

As such, ALFA also supports the development of a Chiral Labeled Release (CLR) instrument concept to measure chiral-specific metabolic activity. However, in an environment where microorganisms are metabolizing extremely slowly or are dormant, mission planners should prioritize the detection of biological phenomena that are detectable without the need to measure active metabolic activity. As mentioned above, molecules and structures conferring function are examples of this. However, if SFL mission planners seek to investigate metabolic activity, the astrobiological payload shall employ an improved version of Viking's Labeled Release (LR) experiment, whereby chiral-specific substrates are applied. The extreme sensitivity, robustness to uncertainties, and widely-applicable nature of the LR experiment informs this position.